\documentclass[preprint,11pt]{elsarticle}
\usepackage{graphicx}
\usepackage{latexsym}

\begin{document}

\title{Matrix Code}

\author{M.H. van Emden}

\maketitle
\begin{abstract}
Matrix Code gives imperative programming a mathematical semantics
and heuristic power comparable in quality to functional and logic
programming.  A program in Matrix Code is developed incrementally
from a specification in pre/post-condition form.  The computations
of a code matrix are characterized by powers of the matrix when it
is interpreted as a transformation in a space of vectors of logical
conditions.  Correctness of a code matrix is expressed in terms of
a fixpoint of the transformation.  The abstract machine for Matrix
Code is the dual-state machine, which we present as a variant of
the classical finite-state machine.

\end{abstract}

\hyphenation{Dijk-stra}

\newtheorem{theorem}{Theorem}{}
\newtheorem{definition}{Definition}{}
\newtheorem{lemma}{Lemma}{}

\newcommand{\emln}{$\;$\\} %empty line
\newcommand{\lmnt}[1]{\parbox{1.0in}{
   {\footnotesize {\emln \tt #1 }} \emln}}
%   {{\emln \tt #1 }} \emln}}
\newcommand{\lmntWdth}[2]{\parbox{#1in}{
   {\footnotesize {\emln \tt #2 }} \emln}}
%    {{\emln \tt #2 }} \emln}}
\newcommand{\nc}[1]{\lmntWdth{0.5}{#1}} % narrow cell
\newcommand{\emp}{\lmnt{$\;$}} %empty cell

\newcommand{\set}[2]{\{#1 \mid #2\}}
\newcommand{\bk}[3]
  {{\tt \{#1\}#2\{#3\}}}
\newcommand{\brr}[2]{{\tt \{#1\}#2}}
\newcommand{\ktt}[2]{#1\{#2\}}

\newcommand{\vc}[3]{${\mathbf V}_{\mbox{{\tt #2}}}(#1;#3)$}
\newcommand{\trpl}[3]{{\tt \{#1\}#2\{#3\}}}
\newcommand{\mpr}[2]{{\tt #1:#2}} % merge pair
\newcommand{\mst}[2]{{\tt (#1,#2)}} % merge state

%pair
\newcommand{\pr}[2]{\ensuremath{\langle #1,#2 \rangle}}

\newcommand{\ccc}{{\tt C}}
\newcommand{\cpp}{{\tt C++}}

\newcommand{\mc}[1]{\mathcal{#1}}

\section{Introduction}
By \emph{imperative programming} we will understand
the writing of code in which the state of the computation
is directly tested
and explicitly manipulated by assignment statements.
As a programming paradigm, imperative programming should be
compared with functional and logic programming.
Compared to these latter paradigms,
imperative programming is in an unsatisfactory state.
At least as a first approximation,
a definition in functional or logic programming
is both a specification and is executable.
In imperative programming proving that a function body
meets its specification is such a challenge
that it is not considered part of a programmer's task.
Another difference, probably related,
is that functional and logic programming
have an elegant mathematical semantics
in which the behaviour of an executable definition
is characterized as a fixpoint of the transformation
associated with the definition.

C is the programming language \emph{par excellence}
for imperative programming.
But in C one can fake functional programming to a certain extent
by doing as much as possible with function definitions,
function calls, and function parameters.
In this paper we will be concerned with what may be called
\emph{hard-core} imperative code:
code in the form of the body of a procedure
({\tt void} function in C) that contains no global variables
and that interacts only with its environment
by reading, testing, and modifying the actual parameters in the call.
These parameters,
together with any local variables that may be present,
comprise the state that is changed by assignment statements.
Surprisingly perhaps, hard-core imperative code does not
exclude function or procedure calls.

In imperative programming verification is a serious problem.
The problem is more serious than in functional or logic
programming because there the executable code can itself be 
the definition of the function or predicate to be executed.
Of course this ideal is rarely reached completely.
But it is a clear ideal for the programmer to strive after.
In imperative programming such an ideal does not exist.
Here correctness has to be proved independently of the code.
Although a powerful verification method was developed
by Floyd and by Hoare,
the experience is that it is hard to produce
a correctness proof for \emph{existing} code.
Dijkstra observed \cite{djk68a,djkInfotech71}
that code has to be \emph{designed
for} correctness proof.
He did not make this suggestion more concrete than
to call for the \emph{parallel development of proof and code}.

This paper is a contribution to the parallel development
of proof and code for imperative programming.
It takes the form of a new language,
called \emph{Matrix Code},
in which programs take the form of a matrix
of which the elements are binary relations among data states.
Matrix Code is distinguished by a development process
that begins with a
null code matrix, progresses with small, obvious steps, and
ends with a matrix that is of a special form that
is trivially translatable to a conventional language
like Java or {\tt C}.
The result of the translation has the same behaviour
as the one determined by the mathematical semantics of the code matrix.
Therefore the latter can be said to be executable.
As every stage in the development process is partially correct
with respect to the specification
(the correctness of the initial null code matrix
is \emph{very} partial).
Matrix Code comes close to the ideal in which
the code is itself a proof of partial correctness.
Matrix Code comes with an abstract machine,
which we call a \emph{dual-state machine} (DSM).
The DSM has easily identifiable special cases that are
trivially translatable to conventional languages like {\tt C} or Java.

\paragraph{Plan of the paper}

Because we derive DSMs from finite-state machines
we first review conventional automata theory
and regular expressions.
The step to DSMs is made by exploiting
the fact that formal languages
are mathematically similar to binary relations
and that both are best regarded as interpretations of
regular expressions.
Accordingly, in Section~\ref{sec:prelim} we establish
our notation and terminology for formal languages, binary
relations, and regular expressions. 

In Section~\ref{sec:FSM} 
we present the main definitions concerning FSMs.
This presentation is necessary because of slight,
yet essential variations in the usual definitions.
One such variation is that the transition is given as a matrix. 
In Section~\ref{sec:DSM} dual-state machines are
introduced as a close variant of FSMs.
The versatility of DSMs is demonstrated
by examples including an FSM, a Turing machine,
and DSMs that translate to {\tt C} programs for generating
prime numbers and for merging files.
As the latter type of DSM is the motivation for the
entire enterprise we devote Section~\ref{sec:MC}
to it.

In Section~\ref{sec:floyd} we
adapt the verification method of Floyd and Hoare to
Matrix Code.
In Section~\ref{sec:sys}
we solve as example problem the generation of prime numbers
in the systematic manner that is specific to Matrix Code.
This is the same problem as one of those treated by Dijkstra
in \cite{djkddh72},
so that Matrix Code can be compared with structured programming.
Although the derivation method for the prime-number
algorithm is original,
the computations of the resulting code are the same
as those of the conventionally produced version.
But Matrix Code is not only valuable as a method
for developing proof and code in parallel,
but, as we show in the derivation of the merging algorithm
in Section~\ref{sec:expressiveness},
it is valuable also
for finding algorithms that are more efficient
than those obtained in the conventional manner.
The final two sections
draw conclusions and survey related work
in widely scattered areas of computer science.

\section{Preliminaries}
\label{sec:prelim}

The dual-state machines to be introduced in this paper
are a variant of the classical finite-state machines.
Just as finite-state machines define formal languages,
dual-state machines define binary relations.
The similarity between the two types of machine
has to do with the similarity between formal languages
and binary relations.
One of the ways this similarity manifests itself
is the fact that formal languages and binary relations
have a natural notation in common: regular expressions.

\subsection{Formal languages}
Given a set $A$, we denote the set of finite sequences
of its elements as $A^*$.
We often think of $A$ as an ``alphabet'',
of its elements as ``symbols'',
of the sequences of symbols as ``words'',
and of sets of words as a (formal) ``language''.

$A^*$ includes the empty word,
the sequence of length 0, which is denoted $e$.
The null language is the empty set.
This is not to be confused with the unit language,
which contains the empty word as its only word.

The concatenation of words $w_0$ and $w_1$ is denoted 
$w_0 \cdot w_1$.
We have
$e\cdot w = w = w\cdot e$
for all words $w$.
Concatenation of words is extended elementwise to
concatenation of languages:
$L_0 \cdot L_1 =
  \{w_0\cdot w_1 \mid w_0 \in L_0 \wedge w_1 \in L_1 \}.$
Concatenation of a language $L$ with itself
gives rise to the powers of $L$:
$L^0$ is the unit language
and $L^i = L\cdot L^{i-1}$ for all $i>0$.
The closure $^*$ of $L$ is defined as
$L^* = \bigcup_{n=0}^\infty L^n$.

The partial order $\leq$ on formal languages is defined
to be set inclusion among the subsets of $A^*$.

\subsection{Binary relations}
A binary relation on a set $D$ is a subset of
the Cartesian product $D \times D$.
If $(d_0,d_1)$ is in a binary relation,
then we say that $d_0$ is an \emph{input};
$d_1$ is a corresponding \emph{output} of the relation.

The null relation is the empty subset of $D \times D$.
The identity relation $I_D$ on $D$
is $\set{(d_0,d_1) \in D \times D}{d_0 = d_1}$. 
The union $R_0 \cup R_1$ of binary relations
$R_0$ and $R_1$  is defined to be
their union as subsets of $D \times D$.
The composition $R_0;R_1$ of binary relations
$R_0$ and $R_1$
is $\set{(d_0,d_1) \in D \times D}
  {\exists d \in D.\; (d_0,d) \in R_0 \wedge (d,d_1) \in R_1} $. 

Powers of a relation $R$ are defined by
$R^n = R;R^{n-1} = R^{n-1};R$ for $n>0$
and $R^n = I_D$ for $n=0$.
We write $R^*$ for $\bigcup_{n=0}^\infty R^n$.

The partial order $\leq$ on binary relations is defined
to be set inclusion among the subsets of $D\times D$.

\subsection{Regular expressions}

The syntax of regular expressions \cite{prr90}
over a given set of constants is
defined as follows.
\begin{enumerate}
\item
The constants, $0$, and $1$ are regular expressions.
\item
If $E$ and $F$ are regular expressions
then so are $E + F$ and
$E\cdot F$.

$nE$ and $E^n$ are shorthand for
\[
\underbrace{E+\ldots+E}_{n \mbox{ {\small times}}}
\mbox{ and }
\underbrace{E\cdot\ldots\cdot E}_{n \mbox{ {\small times}}}
\mbox{ for } n > 0.
\]
$0\cdot E$ is $0$ and $E^0$ is $1$.
\item
If $S$ is a finite set of regular expressions
and $E$ is a regular expression,
then $\Sigma S$ is defined as
$0$ if $S = \emptyset$
and
$
\Sigma (S \cup \{E\})
=
E+\Sigma S
$.
\item
If $E$ is a regular expression,
then so is its closure $E^*$.
\end{enumerate}

In practice a different syntax is used for regular expressions.
We see {\tt EF} for $E\cdot F$,
{\tt E|F} for $E+F$,
{\tt E?} for $0+E$,
and
{\tt E+} for $E\cdot E^*$.

The syntax of regular expressions has several semantics:
algebras of which the elements and operations
can serve as interpretations of regular expressions.
Here these algebras are formal languages and binary relations
and serve as semantics for regular expressions. 
The way we intend these algebras to be semantics for regular
expressions is shown in Figure~\ref{fig:REsem}.

\begin{figure}[htbp]
\begin{center}
\begin{minipage}{4in}
\begin{tabular}{c|c|c|c|c|c}
\parbox{20mm}{regular\\ expressions\\[0mm]}
  & \parbox{10mm}{$0$}
    & \parbox{5mm}{$1$}
      & \parbox{5mm}{$\;+$}
        & \parbox{5mm}{$\;\cdot$}
          & \parbox{5mm}{$\;^*$}
 \\
\hline \hline
\parbox{20mm}{\vspace{2mm}formal\\ languages\\[0mm]}
  & \parbox{20mm}{empty\\ language}
    & \parbox{5mm}{$\;\{e\}$}
      & \parbox{5mm}{$\;\cup$}
        & \parbox{5mm}{$\;\cdot$}
          & \parbox{5mm}{$\;^*$}
 \\
\hline
\parbox{20mm}{\vspace{2mm}binary\\ relations\\[0mm]}
  & \parbox{20mm}{empty subset\\ of $D\times D$}
    & \parbox{5mm}{$\;I_D$}
      & \parbox{5mm}{$\;\cup$}
        & \parbox{5mm}{$\;;$}
          & \parbox{5mm}{$\;^*$}
 \\
\hline
\end{tabular}
\end{minipage}
\end{center}
\caption{\label{fig:REsem}
Formal languages and binary relations as
semantics for regular expressions.
}
\end{figure}

The following equalities in terms of regular expressions
hold for both formal languages and
binary relations as interpretation \cite{cnw71}.
\begin{eqnarray*}
&& E+F = F+E,\; E+E = E,\;
   E\cdot(F+G) = E\cdot F + E\cdot G,\; \\
&& (F+G)\cdot E = F\cdot E + G\cdot E,\; 
   E\cdot(F\cdot G) = (E\cdot F)\cdot G,\;\\
&& 0+E = E+0 = E,\;
   1\cdot E = E \cdot 1 = E,\;
   0\cdot E = E \cdot 0 = 0,\\
&& (E+F)^* = (E^*\cdot F)\cdot E^*,\;
   (E\cdot F)^* = 1+E\cdot(F\cdot E)^*\cdot E,\;\\
&& (E^*)^* = E^*,\;
   E^* = (E^n)^*\cdot E^{<n}
%   \forall i \exists n,j, 0<k<n.
%   E^i = (E^n)^j\cdot E^k 
\end{eqnarray*}
In the last equality $E^{<n}$ denotes the product of
all $k$ such that $0 \leq k < n$.
The axiom states that every power $E^i$ can be written
in the form $(E^n)^j \cdot E^k$ with $0 \leq k < n$.

In addition the partial order $\leq$
defined in the two interpretations for regular expressions
have the property of being
monotonic with respect to the operations.

\section{Finite-state machines}
\label{sec:FSM}

We find the classical finite-state machine useful
as the precursor of the dual-state machine,
a related device with many interesting special cases,
including small imperative programs,
such as are suitable for the bodies of functions in
conventional programming languages.

Our starting point is the nondeterministic
finite-state machine as
a device for recognizing a certain class of languages.
The syntax of finite-state machines is as follows.

\begin{definition}
\label{def:FSMsynt}
A finite-state machine (FSM)
is a tuple $(K,A,\delta,S,H)$
where $K$ and $A$ are nonempty finite sets,
$\delta$ is a function of type
$K \times K \to W$, where $W$ is the set of finite sets of
words over $A$;
$S \in K$, and
$H \in K$.
The elements of $K$ are called \emph{states};
the elements of $A$, the alphabet, are called \emph{symbols};
$\delta$ is the \emph{transition matrix},
a matrix with rows and columns indexed by the elements of $K$.
The transition matrix has sets of words over $A$ as elements.
$S \in K$ is the \emph{start state};
$H \in K$ is the \emph{halt state}.
For all $k \in K$ it is the case that $\delta$
must satisfy $\delta[k,S] = \emptyset$
and $\delta[H,k] = \emptyset$.
\end{definition}

Here we find two departures from the conventional definition:
(1) The transition $\delta$ is a matrix.
In this way we make explicit what is conventionally
left implicit.
(2) There is a single halt state. As is well-known,
this neither adds to nor detracts from the FSM's power.
An advantage of the single halt node is that FSM's
become composable and that one FSM can be substituted into another.

\begin{figure}[htbp]
\hrule
\vspace{0.3cm}
%\begin{minipage}{5.5in}
\begin{center}
\begin{minipage}[b]{2.1in}
\includegraphics[scale = 0.6]{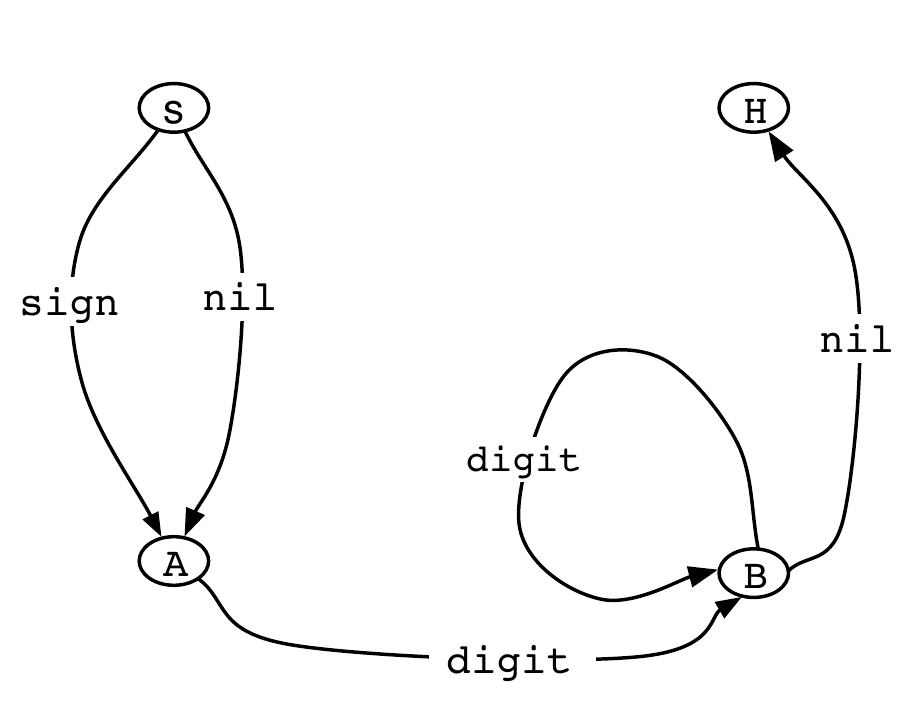}
\vspace{-2cm}
\end{minipage}
\hspace{0.5cm}
\begin{minipage}{1.5in}
{\footnotesize
\begin{tabular}{|l|l|l||l}
\lmntWdth{0.25}{B} & \lmntWdth{0.25}{A} & \lmntWdth{0.25}{S} & \\
\hline \hline
{\tt nil} & & & \lmntWdth{0.21}{H}  \\
\hline
& & {\tt sign} $\cup$ {\tt nil} & \lmntWdth{0.25}{A}  \\
\hline
{\tt digit} &
 {\tt digit} & & \lmntWdth{0.25}{B}  \\
\hline
\end{tabular}
} % footnotesize
\end{minipage}
%\end{minipage}
\end{center}
\caption{\label{fig:FSM}
On the left,
transition matrix in graph form of an FSM that accepts decimal
numerals.
There is no arc from state $k_0$ to state $k_1$
where $\delta[k_0,k_1] = \emptyset$.
The labels on the arcs are
{\tt nil} $= \{e\}$, where $e$ is the empty word,
{\tt sign} $= \{\langle -\rangle,\langle + \rangle\}$
and
{\tt digit} $= \{\langle 0 \rangle,\ldots,\langle 9 \rangle\}.$
Here $\langle x \rangle$ is the word of length 1
containing the symbol $x$.
On the right,
matrix version of Figure~\ref{fig:FSM}.
The rows and columns are identified by the states as labels.
As the row labeled by {\tt S} is by definition empty,
it is omitted.
Similarly, the column labeled by {\tt H} is omitted.
Empty cells are understood to contain $\emptyset$.
}
\vspace{0.2cm}
\hrule
\end{figure}

For an example of an FSM, see Figure~\ref{fig:FSM}.
We choose the FSM to be nondeterministic as starting point
because it is equivalent to some deterministic version
and because it is mathematically more tractable.

Definition~\ref{def:FSMsynt} gives only the syntax of
an FSM.
A common analogy is to view an FSM as a machine
that can do work.
This work is to make computations, as given by the semantics
below.

\begin{definition}
\label{def:FSMbeh}
A \emph{configuration} of an FSM $(K,A,\delta,S,H)$
is a pair $(k,w)$ where $k \in K$
and $w \in A^*$.
A \emph{transition} is a pair
$((k_0,w),(k_1,w\cdot u))$ of configurations
such that $u \in \delta[k_0,k_1]$.
A \emph{computation} of the FSM is a sequence of configurations
such that $(S,e)$ is the first element
and every pair of successive configurations
is a transition of the FSM.

A computation is \emph{complete} if its last configuration
is not the first configuration of any transition.

A complete computation is \emph{successful} if its last
configuration is $(H,w)$ for some $w \in A^*$;
otherwise it is \emph{failed}.

$\mc{L}(K,A,\delta,S,H)$
is the \emph{language accepted by}
the FSM with components $(K,A,\delta,S,H)$;
it is the set of all $w \in A^*$
such there exists a successful
computation
starting in $(S,e)$.
\end{definition}
In the machine analogy of an FSM the transition
$((k_0,w),(k_1,w\cdot u))$ is said to ``consume''
the word $u$ from the input.

\paragraph{Example}
The FSM in Figure~\ref{fig:FSM}
has the following as one of its computations
\begin{center}
\begin{minipage}{3in}
\begin{verbatim}
(S,e)(A,-)(B,-1)(A,-12)(B,-123)(H,-123)
\end{verbatim}
\end{minipage}
\end{center}
The fact that 
\begin{center}
\begin{minipage}{3in}
\begin{verbatim}
(S,e)(A,-)(B,-1)(H,-1)
\end{verbatim}
\end{minipage}
\end{center}
is also a computation shows that this FSM
is nondeterministic.\\[-9mm]
\begin{flushright}$\Box$\end{flushright}

\begin{lemma}
An FSM $(K,A,\delta,S,H)$
has a computation \verb"(x,v),...,(y,v.u)"
of length $n$ iff $u\in \delta^n[x,y]$,
for all $n = 1,2,\ldots$
\end{lemma}
\emph{Proof.} Straightforward induction on $n$.\\[0mm]
\begin{theorem}
$\mc{L}(K,A,\delta,S,H) = \delta^*[S,H].$
\end{theorem}
\paragraph{Proof}
$w \in \mc{L}(K,A,\delta,S,H)$
iff
there exists a computation
$(S,e),\ldots,(H,w)$
iff
there exists an $n$ such that
$w \in \delta^n[S,H]$
iff
$w \in \delta^*[S,H].$\\[-0.9cm]
\begin{flushright}$\Box$\end{flushright}

According to Definition~\ref{def:FSMsynt}
a configuration $(S,w)$ can occur only 
as the first configuration of a computation.
Similarly, a configuration $(H,w)$ can occur only 
as the second of the last configuration of a computation.

The purpose of the computations of an FSM
is to define languages in the form of sets of words
over the alphabet $A$.

An FSM is usually presented as a directed
graph with the states as nodes
and with language $\delta[x,y]$
labeling the arc from state $x$ to state $y$.
We prefer the presentation in Figure~\ref{fig:FSM}.

\section{Dual-state machines}
\label{sec:DSM}
The operation of an FSM consists of two
kinds of changes: a change of state and an advance of the
input when a symbol is accepted.
Imagine giving the machine a data memory in the form
of a file that can be accessed only sequentially.
This memory can take the place of the input.
The accepting of a symbol becomes a change
in the state of the data in memory.
But the elements of $K$,
although not data, are also a kind of memory.
To avoid confusion between these different kinds of memory,
we rename elements of $K$ to \emph{control states};
the states of data memory can then be called \emph{data states}.
A machine with these two kinds of memory
we call a \emph{dual-state machine},
a machine that is very like an FSM:
note the similarity between
Definitions~\ref{def:FSMsynt}
and \ref{def:DSMsyntax};
between
Definitions~\ref{def:FSMbeh}
and \ref{def:DSMbeh}.

The syntax of dual-state machines is defined as follows.
\begin{definition}\label{def:DSMsyntax}
The syntax of a \emph{dual-state machine} (DSM)
is given by a tuple $(K,D,\delta,S,H)$
where $K$ is a nonempty finite set,
$D$ is a nonempty set,
$\delta$ is a function of type
$K \times K \to 2^{D \times D}$,
$S \in K$, and
$H \in K$.
$K$ consists of \emph{control states};
$D$ consists of \emph{data states};
$\delta$ is the \emph{transition matrix},
a matrix with rows and columns labeled by the elements of $K$
with binary relations over $D$ as elements.
$S \in K$ is the \emph{start state};
$H \in K$ is the \emph{halt state}.
For all $k \in K$ it is the case that $\delta$
must satisfy $\delta[k,S] = \emptyset$
and $\delta[H,k] = \emptyset$.
\end{definition}

\subsection{The computations of a DSM}
\label{sec:exec}

The semantics of DSMs is defined as follows.\\[-6mm]
\begin{definition}
\label{def:DSMbeh}
A \emph{configuration} of a DSM $(K,D,\delta,S,H)$
is a pair $(k,d)$ where $k \in K$ is a control state
and $d \in D$ is a data state.

A \emph{transition} is a pair
$((k_0,d_0),(k_1,d_1))$ of configurations
such that $(d_0,d_1) \in \delta[k_0,k_1]$.

A \emph{segment} of the DSM is a sequence of configurations
such that every pair of successive configurations
is a transition of the DSM.
The \emph{length} of a segment is the number of transitions in it.

A \emph{computation} of the DSM is a segment in which
the first configuration is $(S,d)$ for some $d \in D$\/\footnote{
We see that for a configuration $(k,d)$ in a computation
to be followed by $(k',d')$ it is necessary that
$((k,d),(k',d'))$ is a transition.
It is possible that the DSM admits a different transition
$((k,d),(k'',d''))$.
In other words, DSMs are not necessarily deterministic. 
}.

A computation is \emph{complete} if its last configuration
is not the first configuration of any transition.

A complete computation is \emph{successful} if its last
configuration is $(H,d)$ for some $d \in D$;
otherwise it is \emph{failed}.

$\mc{R}(K,D,\delta,S,H)$
is the \emph{relation computed by}
the DSM with components $(K,D,\delta,S,H)$;
it is the set of all $(d,d') \in D\times D$
such there exists a
computation $(S,d),\ldots,(H,d')$.

\end{definition}

It helps to visualize the
Definition~\ref{def:DSMbeh} in the following way
(see Figures~\ref{fig:primes3} and \ref{fig:mrg2}).
Execution of a code matrix consists of 
an execution agent performing a sequence of cycles.
The agent carries a configuration
which is updated during the cycle.
At the beginning of the cycle the agent
carries the configuration $(k,d)$.
It enters the matrix
through the column indexed by $k$
until it encounters a non-empty cell.
Let $r$ be the index of the row in which this cell occurs
and let $R$ be the relation in this cell.
If the data state $d$ of the agent is such that
there is a $(d,w) \in R$,
then the agent exits to the right with configuration
$(r,w)$.
This completes the cycle,
and the agent begins a new cycle
unless it exited through row \verb"H".

The agent may start a cycle in a column that does
not contain a transition having its data state
as input.
In that case the agent does not complete the cycle
and execution fails.

Initially the agent carries
a configuration with control state {\tt S}.
If and when the control state changes to {\tt H},
execution halts with success.

\begin{definition}\label{def:matMult}
Given matrices $M,N \in K\times K \to 2^{D\times D}$
we define their product $M;N$ by
$
(M;N)[i,k] = \bigcup_{j \in K} M[i,j];N[j,k]
$
for all $i,k \in K$.
\end{definition}

Let $I$ be the $K$-labeled matrix of binary relations
over $D$ that has the identity relation on $D$
on the main diagonal and the empty relation elsewhere.
Then we have $I;M = M;I = M$ with $M$ any $K$-labeled matrix
with binary relations over $D$ as elements.
We write $M^n$ for $M^{n-1};M$ for a positive integer $n$
while $M^0 = I$.

We characterize the relation computed by a DSM
in terms of its powers.
First a lemma concerning these powers.

\begin{lemma}\label{thm:dynLaw}
A DSM with transition matrix $\delta$
has a computation containing
a segment $(k,d),\ldots,(k',d')$ of length $n$
iff there exists an $n$ such that $(d,d') \in \delta^n[k,k']$.
\end{lemma}
\emph{Proof}

(If)\\
By induction on the segment length $n$.
If $n=1$ the segment has the form $(k,d),(k',d')$,
so that 
$((k,d),(k',d'))$ is a transition and we have
$(d,d') \in \delta[k,k']$
by the definition of computation.

Induction step.\\
$(d,d'') \in \delta^{n+1}[k,k'']$
implies that there exists an $d'$ and an $k'$ such that
$(d,d') \in \delta^n[k,k']$ and
$(d',d'') \in \delta[k',k'']$.
Hence, by the induction assumption,
there exists a computation segment
$(k,d),\ldots,(k',d')$ of length $n$
and $(d',d'') \in \delta[k',k'']$,
which implies
that there exists a segment
$(k,d),\ldots,(k'',d'')$ of length $n+1$.

(Only if)\\
That $(k_0,d),\ldots,(k_{n-1},d') ,(k_n,d'')$
is a computation implies
(by the induction hypothesis)
that $(d,d') \in \delta^n[k_0,k_{n-1}]$
and $(d',d'') \in \delta[k_{n-1},k_n]$.
By the definition of relational composition this implies that
$(d,d'') \in \delta^n[k_0,k_{n-1}];\delta[k_{n-1},k_n]$.
We have
$$\delta^n[k_0,k_{n-1}];\delta[k_{n-1},k_n]
\subseteq  \bigcup_{j \in K} \delta^n[k_0,j];\delta[j,k_n]
= (\delta^n;\delta)[k',k''] =  \delta^{n+1}[k',k'']
$$
This implies that $(d,d'') \in \delta^{n+1}[k,k'']$.

\begin{theorem}
$\mc{R}(K,D,\delta,S,H) = \delta^*[S,H]$.
\end{theorem}

\emph{Proof}
Suppose that the pair $(d,d')$ of data states
is in the relation computed by $\delta$.
By Definition~\ref{def:DSMbeh}
there exists a computation of $\delta$ that begins with $(S,d)$
and ends with $(H,d')$.
According to Lemma~\ref{thm:dynLaw}
there is an $n$ such that
$(d,d') \in \delta^n[S,H]$.
Hence $(d,d') \in \bigcup_{n=0}^\infty \delta^n[S,H]$.

Suppose that $(d,d') \in \bigcup_{n=0}^\infty \delta^n[S,H]$.
By the finiteness assumptions there exists an $n$ such that
$(d,d') \in \delta^n[S,H]$.
According to Lemma~\ref{thm:dynLaw}
this implies that there exists a computation of $\delta$
that begins with $(S,d)$ and ends with $(H,d')$.
Therefore $(d,d')$ is in the relation computed by $\delta$,
according to Definition~\ref{def:DSMbeh}.

\subsection{Examples of dual-state machines}

\paragraph{Example: FSM as DSM}
A dual-state machine can simulate an FSM.
When doing so, we must keep in mind that the data state
of the DSM is the input of the FSM.
This reverses the role of data in the configurations.
In FSMs the configuration contains the accepted part
of the input, so it grows at transitions.
In the DSM simulation the configuration contains the input of the FSM,
so it shrinks at transitions.

In this example we get a simulation of
the FSM of Figure~\ref{fig:FSM}
by setting $K = \{S,A,B,H\}$,
$D$ equal to the set of words over$\{+,-,0,\ldots,9\}$,
and $\delta$ equal to the matrix in Figure~\ref{fig:FSM}
where {\tt nil} is equal to
$\{(w,w) \mid w \in D\}$,
\verb"digit" is equal to the set of all
$(x\cdot y,y)$ such that $x$ is a word of unit length
over the alphabet $\{0,\ldots,9\}$,
and \verb"sign" is equal to the set of all
$(x\cdot y,y)$ such that $x$ is a word of unit length
over the alphabet $\{-,+\}$.\\[-0.9cm]
\begin{flushright}$\Box$\end{flushright}

\paragraph{Example: Turing Machine as DSM}
We saw that an FSM is a DSM with a certain type of memory
defined by the admissible operations.
A Turing machine is a DSM with a different type of memory
defined by a different set of admissible operations.
Among the several variants of Turing machine
we choose the one where the memory takes the form of
a sequence of squares (a ``tape'') that is unbounded
in both directions.
Each square contains one symbol from the finite alphabet
$A$.
In addition to the contents of the tape,
the state of the memory is determined by a pair $(S,D)$
where $S$ is a square of the tape (the ``scanned'' square) and $D$
is a direction on the tape, being $L$ (left), $R$ (right),
or $d$ (don't care).
The operations on the memory include reading,
a function with no argument having as value
the symbol on the scanned square and writing,
a function with a symbol as argument
causing the scanned square to contain that symbol.
Writing has an additional effect:
to ``move the tape'',
meaning that it causes the scanned square to become
the one on the left or on the right of the currently scanned square,
depending on whether $D$ (``direction'') is $L$ or $R$.

The operation of a Turing machine is determined by a
set of rules, each in the form of quintuple $<Q,S,Q',S',D'>$.
The rule specifies that, if the state is $Q$
and the scanned square contains $S$,
then $S'$ is written, the tape moves in the current
direction $D$, and the state and direction become
$Q'$ and $D'$, respectively.

For this example we selected a simple Turing machine
(\cite{mnsk67}, page 122).

The conventional presentation of the Turing machine
as set of quintuples
is in Figure~\ref{fig:TM}.
The matrix version is in Figure~\ref{fig:TMmtrx}.
The Matrix Code version is in Figure~\ref{fig:TurCode}.
\begin{figure}[htbp]
\begin{center}
\begin{minipage}{4.5in}
\begin{tabular}{l||l|l|l|l|l|l|l|l|l|l|l|l|}
$Q$ &$Q_0$&$Q_0$&$Q_0$&$Q_0$&$Q_1$&$Q_1$&$Q_1$&$Q_1$&$Q_2$&$Q_2$&$Q_2$&$Q_2$\\
\hline
$S$ &$)$&$($&$A$&$X$&$)$&$($&$A$&$X$&$)$&$($&$A$&$X$\\
\hline
$Q'$ &$Q_1$&$Q_0$&$Q_2$&$Q_0$&$Q_1$&$Q_0$&$H$&$Q_1$&$-$&$H$&$H$&$Q_2$ \\
\hline
$S'$ &$X$&$($&$A$&$X$&$)$&$X$&$0$&$X$&$-$&$0$&$1$&$X$\\
\hline
$D'$ &$L$&$R$&$L$&$R$&$L$&$R$&$d$&$L$&$-$&$d$&$d$&$L$
\end{tabular}
\end{minipage}
\end{center}
\caption{\label{fig:TM}
A Turing machine.
The leftmost column shows the generic quintuple
$<Q,S,Q',S',D'>$.
The other twelve columns contain the actual twelve
quintuples that define the Turing machine.
The states are
$Q_0$, $Q_1$, $Q_2$, and $H$.
The tape symbols are ), (, $A$, $X$, $0$, and $1$.
The dashes indicate that in state $Q_2$ the symbol `)'
is never encountered.
The $d$'s in the last row stand for ``don't care''.
}
\end{figure}
As a first step for its simulation by a DSM
we rewrite the conventional Turing machine presentation
to matrix format, which we then find \emph{is} a DSM.
Subsequently we rewrite the code matrix to {\tt C} or {\tt C++}.

\vspace{3mm}

The Turing machine of Figure~\ref{fig:TM}
is designed to start operation with
a tape containing a sequence of parentheses bounded on either
side by the symbol $A$.
Initially the scanned square is
the square containing the leftmost parenthesis,
that is, the square to the right of the leftmost $A$.
When the machine halts,
all matching parentheses have been removed.
In this way one can tell whether the tape initially contained
a well-formed sequence.
Thus, for example,
$$ \mbox{{\tt
A ( ( ( ( ( ( ) ) ) ) A
}}
$$
is replaced by
$$ \mbox{{\tt
A ( 0 X X X X X X X X A, 
}}$$
indicating that the input sequence was unbalanced because
of an unmatched open parenthesis, whereas
$$ \mbox{{\tt
A ( ( ( ( ( ) ) ) ( ) ) ) A 
}}$$
is replaced by
$$ \mbox{{\tt
1 X X X X X X X X X X X X A 
}}$$

The conventional presentation of Turing machines
as a set of quintuples hides their essence,
which is a matrix.
Just as FSMs centre around transitions from state to state,
so do Turing machines.
Whatever the nature of this transition,
its natural presentation is as an element of a matrix
of which the rows and columns are indexed by the states.
Figure~\ref{fig:FSM} gives this matrix for an FSM;
Figure~\ref{fig:TMmtrx} gives this matrix for the Turing
machine in Figure~\ref{fig:TM}.

To familiarize ourselves with the matrix format,
let us find in Figure~\ref{fig:TMmtrx}
the equivalent of the quintuple $<Q_0,),Q_1,X,L>$
of Figure~\ref{fig:TM}.
In this quintuple we see that it specifies
a transition from $Q_0$ to $Q_1$.
The attributes of this transition are in column $Q_0$,
row $Q_1$.
The machine makes this particular transition
if the scanned square contains `)'.
As a result of the transition an $X$ is written on
the scanned square and the tape moves left.
Given that the transition is from $Q_0$ to $Q_1$,
the further particulars can be given
in the form of a condition/action rule:
$$
\mbox{{\tt ) -> X;L}},
$$
which is in the matrix cell of column $Q_0$, row $Q_1$.
Some transitions, for example the one from $Q_0$ to $Q_0$,
contain more than one such rule,
one for each of the possible contents of the scanned square.
See Figure~\ref{fig:TMmtrx}.

\begin{flushright}$\Box$\end{flushright}

\begin{figure}[htbp]
\begin{center}
\begin{minipage}{4in}
\begin{tabular}{|c|c|c||c}
\lmnt{Q2}& \lmnt{Q1} & \nc{Q0} & \\
\hline \hline
\lmnt{( -> 0;d \\ A -> 1;d}
     &\lmnt{A -> 0;d} & \lmnt{}& \nc{H}  \\
\hline
& \lmnt{( -> X;R}  &\lmnt{( -> (;R \\ X -> X;R} & \nc{Q0}\\
\hline
& \lmnt{) -> );L \\ X -> X;L}
      &\lmnt{) -> X;L}  & \nc{Q1}\\
\hline
\lmnt{X -> X;L}
    & & \lmnt{A -> A;L} 
        & \nc{Q2}\\
\hline
\end{tabular}
\end{minipage}
\end{center}
\caption{\label{fig:TMmtrx}
The Turing machine of Figure~\ref{fig:TM} in matrix form;
the initial state is {\tt Q0}.
We propose this as a more readable alternative for the
standard set of quintuples.
}
\end{figure}

\vfill\eject
\section{Matrix Code}
\label{sec:MC}
The main application of DSMs is what we call ``Matrix Code''.
This is a DSM with components $(K,D,\delta,S,H)$ where $D$
is defined by declarations in a conventional programming language,
say $P$, and where the binary relations in the transition matrix
$\delta$ are specified by $P$ with reference to the declarations
for $D$.

To use DSMs to best effect, $P$ should be
equal to \ccc\ or \cpp\
in speed and compactness of compiled code.
Although the binary relations constituting $\delta$
are defined by pieces of code of $P$,
$\delta$ itself is not a construct of $P$ ---
after all, $\delta$ is a \emph{matrix}.
Thus DSMs of this kind constitute a different programming language,
and it is this programming language that we call
\emph{Matrix Code}.
A specific $\delta$ will be referred to as a \emph{code matrix}.

A code matrix is a hybrid object
composed of two programming languages:
Matrix Code and a conventional programming language $P$.
The primitive binary relations of the matrix elements
are written in $P$.
The way they are composed into composite matrix elements
as well as the matrix as a whole are written in Matrix Code.

As we shall show, Matrix Code has two advantages over conventional
languages:
its programs can be their own proof of partial correctness
and it supports
the parallel development of correctness proof and code.
At the same time, a code matrix can be written in such a way as to be
trivially translatable to $P$.
One can say that suitably written code matrices
are ``almost executable''.
For the examples in this paper
we use the following translation method.

The control state is represented by a variable.
Each column is translated to a switch on this variable.
Each cell in the column is then translated
to one of the cases of the switch.
The matrix as a whole is translated
to the body of a {\tt void} function.
The data state of the code matrix becomes
the parameter(s) of the function.

The elements of a code matrix are binary relations
over the data states, as in all DSMs.
In the case of Matrix Code these binary relations are often
composed of primitive relations, which are of two kinds:
guards and statements.
Guards are boolean expressions;
semantically they are subsets of the identity relation.
That is, if $b$ is a boolean expression,
then its meaning is
$\{(d,d) \mid b \mbox{ is true in data state } d\}$.

The primitive relations that are not guards
are statements of $P$.
If $S$ is a statement,
then its meaning is the set of pairs $(d,d')$ such that
$d'$ is a possible state of termination of $S$
if $S$ starts execution in data state $d$.
Because guards and statements both denote
binary relations over $D$,
they are freely intercomposable:
\begin{verbatim}
                 guard ; guard
             statement ; guard
             statement ; statement
                 guard ; statement
\end{verbatim}
are all defined,
and are binary relations over $D$.
As far as Matrix Code is concerned,
\verb"x-- ; x >= 0" and
\verb"x > 0 ; x-- " are equally valid expressions
for binary relations.
The latter form is preferred for reasons
of translatability to a conventional programming language.

\paragraph{Example: code matrix for computing prime numbers}
Consider a DSM with components $(K,D,M,S,H)$
with $K=\{A,B,C,H,S\}$
and $D$ the set of tuples with as components
an integer {\tt N}, an array {\tt p} of length {\tt N},
and integers
{\tt j},
{\tt k}, and
{\tt n}.
$M$ is the code matrix shown in Figure~\ref{fig:primes3}.
For example, $M[C,C]$ is the composition of three binary
relations:\\[-5mm]
\begin{center}
\verb"j%p[n+1] == 0 ; j += 2 ; n = 0"
\end{center}
where \verb"j%p[n+1] == 0" is a boolean expression;
therefore a guard
and \verb"j += 2" and \verb"n = 0" are statements.

See Figure~\ref{fig:compExample} for an example of a computation.

\begin{figure}[htbp]
\hrule \vspace{0.1in}
\begin{center}
\begin{minipage}[t]{3in}
\begin{verbatim}
control | data                     N = 3
state   | state
        | k     j      n    p
-----------------------------------
     S  | ?     ?      ?    {?,?,?}
     A  | 2     ?      ?    {2,3,?}
     B  | 2     5      0    {2,3,?}
     C  | 2     5      0    {2,3,?}
     B  | 2     5      1    {2,3,?}
     A  | 3     5      1    {2,3,5}
     H  | 3     5      1    {2,3,5}
\end{verbatim}
\end{minipage}
\end{center}
\caption{\label{fig:compExample}
Example of the computation for {\tt N} equals 3 of the code matrix
in Figure~\ref{fig:primes3}.
}
\vspace{0.1in}
\hrule
\end{figure}

If a row is empty, then its control state can occur only 
in the first state of any computation.
Such a control state is the start state.
Thus any row indexed by {\tt S} is empty,
and is omitted.
Similarly, the necessarily empty column indexed by {\tt H}
is omitted.
In Figure~\ref{fig:primes3},
$S$ is the start state and
$H$ is the halt state.

\begin{figure}[htbp]
\hrule \vspace{0.1in}
\begin{center}
\begin{minipage}[t]{4in}
{\footnotesize
\begin{verbatim}
int decNum(Tape& t) {
 const int S=0, A=1, B=2, H=3;
 int state = S; // control state
 char inp = t.rd(); // data state component FSM
 int sign, val; // data state component Aux
 while (true) {
  switch (state) {
   case S:
    switch (inp) {
      case'+': sign = +1; inp = t.rd(); break;
      case'-': sign = -1; inp = t.rd(); break;
      default: sign = +1;
    } state = A; break;
   case A: switch (inp){
    case'0':case'1':case'2':case'3':case'4':
    case'5':case'6':case'7':case'8':case'9':
    val = inp - '0';
    inp = t.rd(); break;
    default: assert(false);
    // other character in inp not OK
   } state = B; break;
   case B: switch (inp) {
    case'0':case'1':case'2':case'3':case'4':
    case'5':case'6':case'7':case'8':case'9':
    val = 10*val + (inp-'0');
    inp = t.rd(); state = B; break;
    default: state = H;
   } break;
   case H: return sign*val;
  }
 }
}
\end{verbatim}
} %footnotesize
\end{minipage}
\end{center}
\caption{\label{fig:decNum}
Translation to \cpp\ of
DSM simulation of FSM of Figure~\ref{fig:FSM}.
The data state is in the {\tt Tape\& t}
of the first line.
The object that is the content of the formal parameter
{\tt t} encapsulates the input tape of the FSM.
When the FSM arrives in a state where the first
symbol of the input is accepted,
then the \cpp\ statement {\tt inp = t.rd()}
advances the input tape and makes {\tt inp} again
the first symbol of the remaining input tape.
}
\vspace{0.1in}
\hrule
\end{figure}

\paragraph{Example: code matrix for FSM}
In Figure~\ref{fig:FSM} we presented an FSM for recognizing
a simple form of decimal numerals.
We may regard the input of an FSM, together
with an indication of how far it has been read,
as the data state of a DSM where the control states
are the states of the FSM.
Thus the data state has the form of a file to be read sequentially.

Consider the function \verb"decNum" in Figure~\ref{fig:decNum}.
It is really two programs in one.
If we disregard the line commented with
\verb"data state component Aux" and all assignments to the variables
declared there,
then the remaining part of the data state (\verb"data state component FSM")
is just enough to simulate an FSM:
the variable \verb"inp" contains the first symbol of the
part of the input that has not been processed.
By executing \verb"inp = tape.rd()" the input is advanced by
one symbol,
so that \verb"inp" is once again the first symbol
of the part of the input not processed.

Under the exclusive consideration of
\verb"data state component FSM" the function \verb"decNum"
decides only whether the input is a legitimate decimal number
according to the FSM in Figure~\ref{fig:FSM}.
Usually more is wanted: one might want to know
the value of the decimal numeral.
The advantage of writing the FSM in the form
of a dual-state machine as in Figure~\ref{fig:decNum} is that
one needs to extend only the data state (in this figure to include
\verb"data state component Aux")
and to add appropriate operations
on the extended data state to ensure
that by the time the input is accepted,
the value of the decimal numeral read
is in the data state component \verb"val".

The structure of the function \verb"decNum" reflects the
matrix in Figure~\ref{fig:FSM}.
The outer \verb"switch" translates the matrix column by
column.
Each of the inner \verb"switch" statements translates
the contents of the column concerned by code activated
by the content of \verb"inp".

\begin{figure}[htbp]
\hrule \vspace{0.1in}
\begin{center}
\begin{minipage}[t]{5in}
{\footnotesize
\begin{verbatim}
void turing(Tape& t) {
  typedef enum{H,Q0,Q1,Q2} State;
  State state(Q0); // control state
  while (true) {
    switch (state) {
      case Q0:
        if (t.r() == '(') {t.w('('); t.d(R); state = Q0;} else
        if (t.r() == 'X') {t.w('X'); t.d(R); state = Q0;} else
        if (t.r() == ')') {t.w('X'); t.d(L); state = Q1;} else
        if (t.r() == 'A') {t.w('A'); t.d(L); state = Q2;}
        break;
      case Q1:
        if (t.r() == '(') {t.w('X'); t.d(R); state = Q0;} else
        if (t.r() == ')') {t.w(')'); t.d(L); state = Q1;} else
        if (t.r() == 'X') {t.w('X'); t.d(L); state = Q1;} else
        if (t.r() == 'A') {t.w('0'); t.d(d); state = H;}
        break;
      case Q2:
        if (t.r() == 'X') {t.w('X'); t.d(L); state = Q2;} else
        if (t.r() == ')') {/*can't happen*/ assert(false);} else
        if (t.r() == '(') {t.w('0'); t.d(d); state = H;} else
        if (t.r() == 'A') {t.w('1'); t.d(d); state = H;}
        break;
      case H: return;
      default: assert(false); // can't happen
} } }
\end{verbatim}
} % footnotesize
\end{minipage}
\end{center}
\caption{\label{fig:TurCode}
Code version of the Turing machine in Figures~\ref{fig:TM}
and \ref{fig:TMmtrx}.
The code is in the form of a \cpp\ function with as single
argument {\tt t} an abstract data to represent the tape.
The abstract data type allows three functions:
for reading (called as {\tt t.r}),
for writing (called as {\tt t.w}),
and
for setting tape direction (called as {\tt t.d}).
}

\vspace{0.1in}
\hrule
\end{figure}

\clearpage

\section{Verification of Matrix Code}
\label{sec:floyd}

Verification of Matrix Code is based on Hoare's verification
method for conventional code \cite{hr69},
which in turn is based on Floyd's verification
method for flowcharts \cite{fld67}.
In this section we review Hoare's method,
then show that it can be generalized to binary relations
over any domain,
which in turn gives a verification method for Matrix Code.

\subsection{Hoare's verification method for conventional code}

As an introduction to the verification method
for imperative programming due to Hoare \cite{hr69}
we verify a Java version of the prime-number
generating program
developed by Dijkstra in \cite{djkddh72}.
The Java version of this program is shown in Figure~\ref{fig:floyd}.

\begin{figure}[htbp]
\begin{center}
\hrule \vspace{0.1in}
{\footnotesize
\begin{verbatim}
public static void primes(int[] p, int N) {
  // S
  int j,k,n;
  p[0] = 2; p[1] = 3; k = 2;
  // A
  while (k<N) {
    j = p[k-1]+2; n = 0;
    // B
    while (p[n]*p[n] <= j) {
      // C
      if (j%p[n+1] != 0) n++;
      else {j += 2; n = 0;}
    }
    p[k++] = j;
  }
  // H
}
\end{verbatim}
} % footnotesize
\end{center}
\caption{\label{fig:floyd}
A Java function for filling
{\tt p[0..N-1]} with the first {\tt N} primes.
At the points indicated by the comments S, A, B, C, H
we need conditions to allow verification by Hoare's method.
The identifiers and the structure
are the same as in Dijkstra's example \cite{djkddh72}.
}
\vspace{0.1in}
\hrule
\end{figure}

The concept of configuration consisting
of a control state and data state
used to define the semantics of DSMs applies to conventional
code as well.
Here the control state is a code location
and the data state is the tuple of values of the variables.
According to Hoare's method,
conditions are attached to code locations.
The conditions make assertions about
program variables.
When such a condition occurs in a loop,
it is the familiar invariant of that loop.
In Figure~\ref{fig:floyd} we have indicated by the comments
\verb"S",
\verb"A",
\verb"B",
\verb"C", and
\verb"H"
where these conditions have to be placed. 
Figure~\ref{fig:verification} contains the corresponding
conditions.

\begin{figure}[htbp]
\hrule
\vspace{0.1in}
\begin{center}
\begin{minipage}[t]{3in}
{\footnotesize
\begin{verbatim}
Conditions:
S: p[0..N-1] exists and N>1
H: p[0..N-1] are the first N primes
A: S && p[0..k-1] are the first k primes && k <= N
B: A && k<N && relB(p, k, n, j)
C: B && p[n]*p[n] <= j

relB(p,k,n,j) means that there is no prime
between p[k-1] and j, and that j is not divided
by any prime in p[0..n], and that n<k.

Hoare triples:

{S} p[0]=2; p[1]=3; k=2; {A}
{A && k >= N}  {H}
{A && k < N} j=p[k-1]+2; n=0; {B}
{B && p[n]*p[n] <= j} {C} 
{B && p[n]*p[n] > j} p[k++] = j {A}
{C && j%p[n+1] != 0} n++ {B}
{C && j%p[n+1] == 0} j += 2; n = 0 {B}
\end{verbatim}
} % footnotesize
\end{minipage}
\end{center}
\caption{\label{fig:verification}
Conditions and Hoare triples for Figure~\ref{fig:floyd}.
The meaning of a Hoare triple {\tt \{A0\} CODE \{A1\}}
is that if condition {\tt A0} is true
and if {\tt CODE} is executed
with termination, then condition {\tt A1} is true.
}
\vspace{0.1in}
\hrule
\end{figure}

The verification of the function as a whole
relies on the verification of a number of
implications defined in terms of conditions
and program elements such as tests and statements.
Consider Figure~\ref{fig:floyd}:
because there is an execution path
from \verb+A+ to \verb+B+, one has to show
the truth of\\
\verb"        {A && k<N} j=p[k-1]+2; n=0; {B},"\\
which has as meaning:
if \verb"A && k<N" (the \emph{precondition}) is true
and if\\
\verb"        j=p[k-1]+2; n=0;"\\
is executed, then \verb"B" (the \emph{postcondition})
is true.
Because of the three elements: precondition,
postcondition, and the item in between,
this is called a \emph{Hoare triple}.
Figure~\ref{fig:verification} contains not only
the conditions for Figure~\ref{fig:floyd},
but also the set of verification conditions
in the form of Hoare triples.

The term ``condition'' for the type of thing
that occurs as precondition and postcondition
in a Hoare triple is, in our view, rather compelling.
However, it seems that in certain contexts
``assertion'' is a more natural alternative term.
In this paper we will use both.
At the same time, one should make a distinction
between the condition as a linguistic expression
and the set that is the meaning of that expression.
We trust no confusion arises as we
use ``assertion'' and ``condition''
interchangeably for both the expression and the meaning.

According to Hoare's method
the program in Figure~\ref{fig:floyd}
is verified by the truth of the Hoare triples in
Figure~\ref{fig:verification}.
Why the set of conditions used is necessary and sufficient
and how partial correctness follows from the truth of the
Hoare triples requires a non-negligible amount of explanation.
This can be omitted here because in the following we give
the equivalent explanation for Matrix Code,
for which it will be comparatively simple.

\subsection{Binary relations, conditions, and Hoare triples}

Let us consider a binary relation $R$,
a subset of $D\times D$,
where we can think of $D$ as a set of data states.
Let us call subsets of $D$ \emph{conditions}.
The \emph{left projection} of $R$
is defined as the condition
$\set{x \in D}{\exists y \in D.\; (x,y) \in R}$.
Dually, the \emph{right projection} of a binary relation $R$
is defined as the condition
$\set{y \in D}{\exists x \in D.\; (x,y) \in R}$.

We generalize $I_D$ to $I_c$,
which means, for any condition $c \subseteq D$, by definition,
$\set{(x,x) \in D \times D}{x \in c}$.
This induces a one-to-one relation between $c$ and $I_c$:
$$
x \in c \leftrightarrow (x,x) \in I_c.
$$
Accordingly, at times we view a condition (alias \emph{assertion})
as a subset of $D$; at times as a subset of $I_D$.

\begin{definition}\label{def:projections}
Given a condition $c \subseteq D$
and a binary relation $R \subseteq (D \times D)$,
we write \brr{c}{R}\ for
the right projection of $I_c;R$,
where $I_c$ is the binary relation
$\set{(x,x) \in D \times D}{x \in c}$.
\end{definition}

As we saw above,
Hoare triples were intended to be applied to program statements.
Here we see that
they have a natural interpretation for binary relations.

\begin{definition}\label{def:triple}
Given conditions
$p \subseteq D$ and $q \subseteq D$
and a binary relation $R \subseteq (D \times D)$,
we define that \bk{p}{R}{q} (\emph{the Hoare triple})
holds iff
\begin{center}\brr{p}{R} $\subseteq$ {\tt q}.\end{center}
\end{definition}
That is, if the input to $R$ satisfies {\tt p},
then all corresponding outputs (if any) satisfy {\tt q}.

We extend Definition~\ref{def:projections}
to vectors and matrices.
\begin{definition}\label{def:projVec}
Let $v$ be a vector of conditions: $v \in K \to 2^D$
and let $M$ be a matrix of binary relations:
$M \in K\times K \to 2^{D\times D}$.
Then $\{v\}M$ is defined to be the vector in $K \to 2^D$
such that
%$(\brr{v}{M})[i] = \bigcup_{j\in K}\brr{v[j]}{M[j,i]}$.  
$(\{v\}M)[i] = \bigcup_{j\in K}\{v[j]\}M[j,i]$.  
\end{definition}

\begin{definition}
\label{def:braKet}
Let $p,q \in K \to 2^D$ be vectors of conditions
indexed by $K$
and $M \in K\times K \to 2^{D\times D}$
a matching matrix of binary relations over $D$.
The expression $\{p\}M\{q\}$
asserts that
$(\{p\}M) \subseteq q$
where the set inclusion is taken elementwise.
\end{definition}

\begin{theorem}\label{thm:Floyd}
Given a code matrix $M$
and a condition vector $V$ satisfying
$\{V\}M\{V\}$.
For any configuration $(k',d')$ of any computation beginning with
$(k,d)$ such that $d \in V[k]$ it is the case that
$d' \in V[k']$. 
\end{theorem}
\emph{Proof}\\
We proceed by induction on the length $n$ of the computation.
If $n=1$ (one transition in the computation) we have $(k',d')=(k,d)$.
Assume the theorem true for computations of length $n-1$.
Consider the computation
$$
(k,d),
(k_1,d_1),
\ldots,
(k_{n-1},d_{n-1}),
(k',d').
$$
By the induction assumption $d_{n-1} \in V[k_{n-1}]$. 
We have that $(d_{n-1},d') \in M[k_{n-1},k']$.
It is given that $\bk{V}{M}{V}$,
hence in particular that
$$\{V[k_{n-1}]\}M[k_{n-1},k']\{V[k']\}.$$
It follows that $d' \in V[k']$, which establishes
the theorem for the computation of length $n$.

\section{Parallel development of proof and code}
\label{sec:sys}

Floyd's method is difficult to apply
because it is difficult to find the required conditions.
Because of this Dijkstra \cite{djk68a,djkInfotech71}
advocated parallel development of code and proof.
In this section we demonstrate parallel development
of a code matrix for the sample problem
solved in Figure~\ref{fig:floyd}:
to fill an array with the first $N$ prime numbers
in increasing order.

\paragraph{Background on prime numbers}
Before we start,
let us review what we need to know about prime numbers.
The following list of facts is not intended as a
complete or nonredundant set of axioms;
they are a selection to guide us in the choice
of conditions and transitions.
\begin{enumerate}
\item
\emph{A prime is a positive integer that has no divisors.}
(We do not count 1 or the integer itself as divisors.
Moreover, 1 is not a prime.)
\item \label{axiom:infinity}
\emph{There are infinitely many primes},
so the problem can be solved for any $N$.
\item \emph{2 and 3 are the first two primes}.
So a way to get started is to accept these as given
and place them in the beginning of the table.
This has the advantage
that we always have the situation
where the last prime in the table is odd
and the next odd number is the first candidate to be tested
for the next prime.
\item \label{axiom:suff}
\emph{If a number has a divisor,
then it has a prime divisor.}
This can be used to save effort:
we have to test only for divisibility by smaller primes,
and these are already in the table.
\item \label{axiom:limit}
\emph{If a number has a divisor,
then it has a prime divisor
less than or equal to its square root.}
This implies that we do not have to test
the candidate for the next prime for divisibility
by all primes already in the table.
\item \label{axiom:square}
\emph{The square of every prime is greater than the next prime.}
The significance of this fact will become apparent as we proceed.
\end{enumerate}

\paragraph{Deriving the code matrix}
The distinctive advantage of Matrix Code is
that a matrix can be expanded from the specification
in small steps using only the \emph{logic}
of the application without needing to attend
to the \emph{control} component of the algorithm.
Thus Matrix Code is an example of Kowalski's principle
``Algorithm = Logic + Control'' \cite{kwl79a}.

We assume that the specification exists
in the form of a precondition and a postcondition.
This gives rise to code matrix with one row and one column;
the one in Figure~\ref{fig:primes0}.

\begin{figure}[htbp]
\begin{center}
\begin{tabular}{|l||l}
\lmnt{S: p[0..N-1] exists \& N>1} & \\
\hline \hline
        \lmnt{/*which T?*/}
        & \lmnt{H: p[0..N-1] contains the first N primes}  \\
\hline

\end{tabular}
\end{center}
\caption{\label{fig:primes0}
There is only an empty transition $T$ such that $\bk{S}{T}{H}$.
}
\end{figure}

The one element of this matrix is the transition \verb"T"
such that \bk{S}{T}{H} is true.
That is, \verb"T" has to be a simple combination
of guards and assignment statements that places
the \verb"N" first primes in \verb"p",
whatever \verb"N" is.
Absent such a \verb"T", we leave the matrix cell empty.
The resulting code matrix
satisfies \bk{S}{T}{H}, which makes it partially correct,
but \emph{very} partially so:
it has no successful computations.
Although Figure~\ref{fig:primes0} is the correct start
of the development process, it is not the last step.

As it is too ambitious to place all primes in the array
with a single transition,
a reasonable thing to try is to fill it with the first \verb"k"
primes and then try to add the next prime after \verb"p[k-1]".

We need a condition {\tt A} that is intermediate in the sense
that \bk{S}{T1}{A} and \bk{A}{T2}{H}
for simple {\tt T1} and {\tt T2}.
Such a condition is:
the first {\tt k} primes in increasing order are in
{\tt p[0..k-1]} with {\tt 1 < k <= N}.

Condition \verb"A" is promising because it is easy to think
of such a
\verb"T1"
and such a
\verb"T2".
The result is in Figure~\ref{fig:primes1}.

\begin{figure}[htbp]
\begin{center}
\begin{minipage}{3in}
\begin{tabular}{|l|l||l}
\lmnt{A:} & \lmnt{S: p[0..N-1] exists \& N>1} & \\
\hline \hline
\lmnt{k >= N} & & \lmnt{H: p[0..N-1] contains the first N primes}  \\
\hline
& \lmnt{p[0] = 2; p[1] = 3; k = 2}
   & \lmnt{A: p[0..k-1] contains the first k primes \&
k <= N}  \\
\hline
\end{tabular}
\end{minipage}
\end{center}
\caption{\label{fig:primes1}
In column $A$ the case {\tt k < N} is missing.
}
\end{figure}

This again is a partially correct code matrix.
It is a slight improvement in that it solves the problem
if $N$ happens to be one or two.
In all other cases it leads to failed computations. 
The difficulty is that in column {\tt A}
we may have that {\tt k < N},
so that we cannot make the transition to {\tt H}.
We need to find the next prime after {\tt p[k-1]}.
Let {\tt j} be the current candidate for this next
prime.
That suggests for condition {\tt B:}
{\tt A} is true and {\tt j} is such that
there is no prime greater than {\tt p[k-1]}
and less than {\tt j}.

This is always true when {\tt j} is the next odd
number after {\tt p[k-1]}.
Another way of saying this is that {\tt j} is not
divisible by any of the primes in {\tt p[0..n]} with {\tt n}
set to 0.
We are interested more generally in
\begin{quote}
There are no primes between {\tt p[k-1]} and {\tt j} (with
{\tt j} is not divisible by any of the primes in {\tt p[0..n]})
and {\tt n<k}.
\end{quote}
We abbreviate this condition to \verb"relB(p,k,n,j)".

The largest prime factor of a number is less than the square root
of the number.
Hence, if we find that the square of {\tt p[n+1]} is greater
than {\tt j}, then we can conclude that {\tt j}
is the next prime after \verb"p[k-1]".
Hence, in the new column \verb"B", it is easy to detect
whether \verb"n" is large enough to conclude that \verb"j"
is the next prime after \verb"p[k-1]".
We place the corresponding transition in column \verb"B"
and we have Figure~\ref{fig:primes2}.

\begin{figure}[htbp]
\begin{center}
\begin{minipage}{5in}
\begin{tabular}{|l|l|l||l}
\lmnt{B:} & \lmnt{A:} & \lmnt{S: p[0..N-1] exists \& N>1} & \\
\hline \hline
& \lmnt{k >= N} & & \lmnt{H: p[0..N-1] contains the first N primes}  \\
\hline
\lmnt{p[n]*p[n]>j; p[k++]=j} &
  & \lmnt{p[0] = 2; p[1] = 3; k = 2}
   & \lmnt{A: p[0..k-1] contains the first k primes \&
k <= N}  \\
\hline
        &\lmnt{k<N; j = p[k-1]+2; n=0}  &
           & \lmnt{B: A \& k<N \& relB(p,k,n,j)}                       \\
\hline
\end{tabular}
\end{minipage}
\end{center}
\caption{\label{fig:primes2}
In column $A$ we have added a transition in column $A$
for the case that {\tt k < N}.
In that case we can start finding the next prime after
{\tt p[k-1]} because we know that there is enough space
in {\tt p} to store it.
{\tt relB(p,k,n,j)} means that
there is no prime between the last prime found and {\tt j}
and that {\tt n<k}, and that {\tt j} is not divided
by any prime in {\tt p[0..n]}.
}
\end{figure}

There are still failed computations.
(In fact, there is still no way to get beyond $N = 2$.)
The way ahead is clear:
a transition is missing in column \verb"B",
for the situation where \verb"n" is too small to
conclude that \verb"j" is the next prime.
That in itself produces condition \verb"C"
and, with it, a new row and column.

In column \verb"C" the missing information
is whether \verb"j",
the candidate for the next prime,
is divisible by \verb"p[n+1]".
If not, then \verb"n" can be incremented,
and condition \verb"B" is verified.
If so, then \verb"j" is not a prime
and the search for the next prime
must be restarted with \verb"j+2".
This determines a transition in column \verb"C"
that verifies condition \verb"C",
so is placed in that row.
See Figure~\ref{fig:primes3}.

\begin{figure}[htbp]
\begin{center}
\begin{minipage}{5.5in}
%???
%\setlength{\leftmargin}{-0.5in}
\begin{tabular}{|l|l|l|l||l}
\lmnt{C:}& \lmnt{B:} & \lmnt{A:}
     & \lmnt{S: p[0..N-1] exists \& N>1} & \\
\hline \hline
& & \lmnt{k >= N} & & \lmnt{H: p[0..N-1] contains the first N primes}  \\
\hline
& \lmnt{p[n]*p[n]>j; p[k++]=j} &
  & \lmnt{p[0] = 2; p[1] = 3; k = 2}
   & \lmnt{A: p[0..k-1] contains the first k primes \&
k <= N}  \\
\hline
\lmnt{j\%p[n+1]!=0; n++} & &\lmnt{k<N; j = p[k-1]+2; n=0}  &
           & \lmnt{B: A \& k<N \& relB(p,k,n,j)}                       \\
\hline
\lmnt{j\%p[n+1]==0; j += 2; n=0}
    & \lmnt{p[n]*p[n]<= j} &
       & & \lmnt{C: B \& \\ p[n]*p[n] <= j}\\
\hline
\end{tabular}
\end{minipage}
\end{center}
\caption{\label{fig:primes3}
This figure is both a general example of a code matrix
and the final stage of the development
consisting of the sequence of Figures
\ref{fig:primes0},
\ref{fig:primes1}, and
\ref{fig:primes2}.
Change from Figure~\ref{fig:primes2}:
row and column with label $C$ are added.
There are no incomplete columns.
This, as well as each of the previous versions is
partially correct,
as implied by the validity of
the verification condition for each
of the null matrix elements.
The absence of incomplete columns opens the possibility
of total correctness, but does not prove it.
}
\end{figure}

Up till now we detected with every additional
row and column that the new column lacked a transition.
Not this time: none of the columns has a missing transition.
The code matrix has no failed computations.
So it gives the correct answer by exiting in row \verb"H",
or it continues in an infinite computation.
As we have proved only partial correctness,
this latter alternative remains a possibility.

\paragraph{Termination}
For an infinite computation to arise,
there must be at least one condition
that is revisited an infinite number of times.
For each condition we give a reason why
it can be revisited only a finite number of times.

\begin{enumerate}
\item
Condition {\tt A.}
For this condition to be returned to,
{\tt k} has to have increased.
{\tt k} is never decreased and is bounded by {\tt N}.

\item
Condition {\tt B.}
For this condition to be returned to,
{\tt n} or {\tt j} has to have increased.
{\tt n} is bounded by the square root of 
{\tt p[N-1]}.
The number of times it is reset to zero is bounded by
{\tt p[N-1]}.
{\tt j} is never decreased and is bounded by {\tt p[N-1]}.
\item
Condition {\tt C.}
For this condition to be returned to,
{\tt n} has to have increased and is bounded as noted above.
\end{enumerate}

The transitions have been chosen
so that the corresponding revisiting condition
is satisfied.
As none of these conditions can be satisfied
an infinite number of times,
the code matrix has no infinite computation.

\paragraph{Running Matrix Code}
Running a code matrix in current practice
requires translation to a currently available language.
Our examples of Matrix Code have been constructed
for ease of translation to languages like Java or {\tt C}.
This entails a drastic reduction in expressivity.
Let us now demonstrate translation
using Figure~\ref{fig:primes3} as example.

As there is a similarity between the control states
and the states of a finite-state machine (FSM),
a good starting point for systematic translation
of a code matrix is the pattern according to which
an FSM is implemented.
This is usually done by introducing a constant for every
state and to let a variable, say, \verb"state"
assume these constants as values.
An infinite loop containing a \verb"switch" controlled by
\verb"state" then contains a \verb"case" statement
for every control state.
%The fact that in a programming language the case statements
%are not restricted to input or output
%is the generalization that produces a code matrix from
%an FSM.

Each column of a code matrix translates
to a \verb"case" statement.
The order in which the translations of the columns
occur does not matter as long as \verb"state"
is initialized at \verb"S".
Here we have arbitrarily chosen alphabetic order.
In this way Figure~\ref{fig:primes3} translates to
the following.

\begin{figure}[htbp]
\hrule \vspace{0.1in}
\begin{center}
\begin{minipage}{3in}
{\footnotesize
\begin{verbatim}
void prTable(int p[], int N) {
  typedef enum{A,B,C,H,S} State;
  State state(S); // control state
  int j,k,n;      // part of data state
  while (true) {
    switch(state) {
      case A:
        if(k >= N) state = H;
        else {j = p[k-1]+2; n = 0; state = B;}
        break;
      case B: if (p[n]*p[n] > j) {
                p[k++] = j; state = A;
              } else state = C;
              break;
      case C:
        if (j%p[n+1] != 0) {n++; state = B;}
        else {j += 2; n = 0; state = C;}
        break;
      case H: return;
      case S: p[0] = 2; p[1] = 3; k = 2; state = A;
    }
  }
}
\end{verbatim}
} % footnotesize
\end{minipage}
\end{center}
\caption{\label{fig:primesCM}
Translation of the code matrix
in Figure~\ref{fig:primes3} to \cpp.
}
\vspace{0.1in}
\hrule
\end{figure}
A transition {\tt b0;S0} in column $X$ and row $R_0$ and
transition {\tt !b0;S1} in column $X$ and row $R_1$
translate to
{\tt case X: if (b0) \{S0; state = R0;\}
else \{S1; state = R1\} break;} in the above code.

\section{Expressiveness of Matrix Code}
\label{sec:expressiveness}

The code obtained by translating a code matrix
is quite different from what one conventionally would write:
compare Figure~\ref{fig:floyd} with
Figure~\ref{fig:primesCM}.
In this example Matrix Code has the advantage of being
a verification and of being easy to discover.
But in the prime-number problem
Matrix Code does not lead to a more efficient program:
it has the same set of computations as the conventional one.

In this section we present an example where Matrix Code
makes it easy to discover an algorithm that is
more efficient than what is obtained via the conventional
programming style.
Consider the merging of two
monotonically nondecreasing input streams
into a single output stream.
We have available the following \cpp\ functions.

\begin{verbatim}
bool getL(int& x);  // output parameter x
bool getR(int& x);  // output parameter x
void putL();
void putR();
\end{verbatim}
where {\tt getL} ({\tt getR})
tests the left (right) input stream for emptiness.
In case of nonemptiness the output parameter
{\tt x} gets the value of the first element
of the stream.
Neither {\tt getL} nor {\tt getR} change
any of the streams.
This is done only by the functions
{\tt putL()} and
{\tt putR()}
which transfer the first element of a nonempty
left or right input stream to the output stream.

Figure~\ref{fig:eMerge}
is a typical program for this situation.
It typically acts in two stages.
In the first stage both input streams are nonempty.
In the second stage one of the input streams is empty
so that all that remains to be done
is to copy the other stream to the output.

\begin{figure}[htbp]
\hrule \vspace{0.1in}
\begin{center}
\begin{minipage}[t]{2.5in}
\begin{verbatim}
void eMerge() {
  int u,v;
  while (getL(u) && getR(v))
    if (u <= v) putL();
    else        putR();
  while (getL(u)) putL();
  while (getR(v)) putR();
}
\end{verbatim}
\end{minipage}
\end{center}
\caption{\label{fig:eMerge}
A structured program for merging two streams.
}
\vspace{0.1in}
\hrule
\end{figure}

This algorithm performs unnecessary tests:
in the first stage only one of the
input streams is changed, so that only that
one needs to be tested for emptiness;
here both are tested\footnote{
With the one exception when the left input
stream runs out at the same time as, or before,
the right input stream.
}.
It is superfluous tests like this
that allow the algorithm to be as simple as it is.

Of course it is unlikely that it is important
to save the kind of test just mentioned.
But there are many types of merging situations
and there may be some in which it does matter.
An advantage of Matrix Code is that 
it does not bias the programmer
towards including superfluous tests.

We proceed to develop a code matrix for merging.
The assertions need to indicate whether
it is known that an input stream is empty
and, if not, what its first element is.
If an input stream is possibly empty
then we represent it by ``{\tt ?}''.
We write ``{\tt e}'' if an input stream is empty.
Nonemptiness is indicated by writing ``\mpr{x}{?}'',
where {\tt x} is the first element.
We have to do this for each of the input streams;
we write e.g. the assertion \mst{\mpr{u}{?}}{\mpr{v}{?}}
to mean that both input streams are nonempty
and have first elements {\tt u} and {\tt v}, respectively.

We write all conditions in the form {\tt (left,right)},
where {\tt left} and {\tt right}
indicate the state of the input concerned,
in conjunction with the statement
that the result of appending the output
to the result of merging the remaining input streams
is equal to the result of merging the input streams
before the beginning of the execution of the program.
As this conjunct is part of every condition,
it need not be stated explicitly.
Of course its validity needs to be verified for every matrix entry.

With these conventions
we can state the program's specification
as obtaining a transition from
the state {\tt S}, which is {\tt (?,?)}
to
the state {\tt H}, which is {\tt (e,e)}.
%
%the state {\tt H}, which is {\tt (e,e)},
%while maintaining the invariant that
%the result of appending the output stream to the result
%of merging the input streams is constant.
Accordingly, the development starts with Figure~\ref{fig:mrg0}.

\begin{figure}[htbp]
\begin{center}
\begin{tabular}{|l||l}
          \lmnt{S:\mst{?}{?}} & \\
\hline \hline
        \lmnt{/*which T?*/}
        & \lmnt{H:\mst{e}{e}}  \\
\hline \\

\end{tabular}
\end{center}
\caption{\label{fig:mrg0}
Matrix Code corresponding to specification of the merging program.
But there is no {\tt T} such that \bk{S}{T}{H}.
The conditions in this figure,
as well as those in Figures
\ref{fig:mrg1}
and
\ref{fig:mrg2}
include the unstated
conjunct that the result of appending the output stream to
the merge of the input streams is equal to the merge of the input
streams in the initial state.
}
\end{figure}
As always with Matrix Code,
we start with the conditions.
Which do we need, in addition to the
\mst{?}{?}
and
\mst{e}{e}
given by the specification?
For each of the input streams
there are three states of information:

\begin{itemize}
\item
{\tt ?}
%\vspace{-0.3cm}
\item
{\tt e}
%\vspace{-0.3cm}
\item
\mpr{x}{?} for some first element {\tt x}
\end{itemize}
It is to be expected that the two input streams
can assume each of the three information states
independently, for a total of nine conditions. 

It is desirable that the initial condition \mst{?}{?} of 
minimal information does not arise during a computation
of the code matrix.
Under the assumption that we can avoid this
there will be only rows for the eight other conditions. 
By the time we will have populated the columns for these eight
conditions we will see whether this assumption was justified.

This problem is easy because the conditions are determined
by the nature of the problem.
For each condition there is an obvious and easy-to-realize
revisiting condition.
If there is at least one unknown input stream
at least one of them has to become known before revisiting.
If both input streams are known,
then at least one of them has to have its first element
transferred to output before revisiting.
See Figures~\ref{fig:mrg1} and \ref{fig:mrg2},
where the transitions have been chosen to conform
to the revisiting requirements.
As each column either has no guard or two complementary guards,
no additional rows are needed.

\begin{figure}[htbp]
\begin{center}
\begin{tabular}{|l|l||l}
\lmnt{{\tt A}}  &
          \lmnt{S:\mst{?}{?}} & \\
\hline \hline
     &      & \lmnt{H:\mst{e}{e}}  \\
\hline
     &\lmnt{{\tt getL(u)}}&\lmnt{{\tt A:(u:?,?)} }
\\
\hline
     &\lmnt{{\tt !getL(u)}}&\lmnt{{\tt B:(e,?)}}
\\
\hline
     \lmnt{{\tt getR(v)}}&&\lmnt{{\tt C:(u:?,v:?)}}
\\
\hline
     \lmnt{{\tt !getR(v)}}&&\lmnt{{\tt D:(u:?,e)}}
\\
\hline

\end{tabular}
\end{center}
\caption{\label{fig:mrg1}
See Figure~\ref{fig:mrg0}.
An input stream needs to be tested;
the left one is chosen arbitrarily.
This gives rise to new conditions.
Columns for these will cause addition of yet more conditions.
See Figure~\ref{fig:mrg2}.
}
\end{figure}

\begin{figure}[htbp]
\begin{center}
\begin{minipage}{6.5in}
\begin{tabular}{|l|l|l|l|l|l|l|l||l}
\nc{G} & \nc{F} & \nc{E} & \nc{D} & \nc{C} &
\nc{B} & \nc{A} & \nc{S:(?,?)} & \\
\hline \hline
& \nc{!getL(u)} &  &  &  &
\nc{!getR(v)} &  &  & \nc{H:(e,e)} \\
\hline
& & & & \nc{u>v;\\ putR()} &
&  & \nc{getL(u)} & \nc{A:(u:?,?)}\\
\hline
\nc{putR()} &  &  &  & &
\nc{getR(v);\\putR()} & & \nc{!getL(u)} & \nc{B:(e,?)}\\
\hline
& & \nc{getL(u)} & & & & \nc{getR(v)} &  & \nc{C:(u:?,v:?)}\\
\hline
& \nc{getL(u)} & & &  & & \nc{!getR(v)} &  & \nc{D:(u:?,e)}\\
\hline
 &  &  &  & \nc{u <= v;\\ putL()} & &  &  & \nc{E:(?,v:?)}\\
\hline
 &  &  & \nc{putL()} &  & &  &  & \nc{F:(?,e)}\\
\hline
 &  & \nc{!getL(u)} &  &  & &  &  & \nc{G:(e,v:?)}\\
\hline
\end{tabular}
\end{minipage}
\end{center}
\caption{\label{fig:mrg2}
The complete code matrix for the merging problem,
continuing Figures \ref{fig:mrg0} and \ref{fig:mrg1}.
}
\end{figure}

The translation of the code matrix in
Figure~\ref{fig:mrg2}
is given in
Figure~\ref{fig:mMerge}.
As the order of the translations of the columns
is immaterial,
we have placed them in alphabetic order by label.

\begin{figure}[htbp]
\hrule \vspace{0.1in}
\begin{center}
\begin{minipage}{4in}
{\footnotesize
\begin{verbatim}
void mMerge(Trinity& tri) {
  int u,v;
  typedef enum{S,A,B,C,D,E,F,G,H} State;
  State state = S; // control state
  while(true) {
    switch(state) {
      case A: state = (tri.getR(v))?C:D; break;
      case B: if (tri.getR(v)) {tri.putR(); state = B;}
         else state = H; break;
      case C: if (u <= v) {tri.putL(); state = E;}
         else {tri.putR(); state = A;} break;
      case D: tri.putL(); state = F; break;
      case E: state = tri.getL(u)?C:G; break;
      case F: state = tri.getL(u)?D:H; break;
      case G: tri.putR(); state = B; break;
      case H: return;
      case S: state = tri.getL(u)?A:B; break;
    }
  }
}
\end{verbatim}
} % footnotesize
\end{minipage}
\end{center}
\caption{\label{fig:mMerge}
A \cpp\ function for merging two streams
translated from Figure~\ref{fig:mrg2}.
{\tt tri} is an object of class {\tt Trinity}.
It contains three components: two input streams and an output stream. 
The admissible operations on these components are
{\tt getL(u)}, which is to determine the value
of the first element from the
left input stream (if there is one) and to make the argument
{\tt u} equal to it. The input stream is left unchanged.
{\tt putL()} removes the first element from the left input
stream and makes it the next element of the output stream.
Similarly for {\tt getR(v)} and {\tt putR()}
for the right input stream.
}
\vspace{0.1in}
\hrule
\end{figure}

The reason for developing a code matrix for the merge problem
was the desire to avoid the superfluous tests of a function
like the {\tt eMerge} listed in Figure~\ref{fig:eMerge}.
To see in how far {\tt mMerge} improves in this respect
we have run both functions on the same set of pairs of input
streams and counted the calls executed in both merge functions.

Such comparisons are of course dependent on the nature of the
input streams.
For example, the more equal in length the input streams are,
the more favourable for {\tt mMerge}.
Accordingly we have used a random-number generator
to determine the lengths of the input streams.
The input streams themselves are monotonically
increasing with random increments.

\begin{center}
\begin{tabular}{l||r|r|r|r}
    & getL & getR & putL & putR \\
\hline \hline
eMerge  & 1756 & 2691 & 871 & 1819 \\
\hline
mMerge  & 872 & 1821 & 871 & 1819 \\
\hline \hline
eMerge  & 1067 & 830 & 655 & 410 \\
\hline
mMerge  & 656 & 411 & 655 & 410 \\
\hline \hline
eMerge  & 3261 & 735 & 2894 & 365 \\
\hline
mMerge  & 2895 & 366 & 2894 & 365 \\
\hline \hline
eMerge  & 1355 & 1024 & 844 & 509 \\
\hline
mMerge  & 845 & 510 & 844 & 509 \\
\hline \hline
\end{tabular}
\end{center}

Each pair of successive lines gives the result of
running {\tt eMerge} and {\tt mMerge}
on the same pair of input streams.
The lengths of the streams are not listed separately,
as they are equal to the number of calls to {\tt putL}
and {\tt putR} shown in the table.

A merge function needs to make at least one call
to {\tt getL} ({\tt getR}) for every element of the left (right)
input stream.
It can be seen that {\tt mMerge} remains close to this minimum,
while {\tt eMerge} does not.

This example is notable
in that Matrix Code yields
an unfamiliar, test-optimal algorithm
by \emph{default}.
Structured programming tends to reduce
the number of control states.
Matrix Code lacks this bias:
in its use it is natural to introduce control states
as needed to serve as memory for test outcomes.

%The raw material for the tables is the following
%computer output:
%
%\begin{verbatim}
%eMerge 54321
%getL: 1756 getR: 2691
%putL: 871 putR: 1819
%mMerge 54321
%getL: 872 getR: 1821
%putL: 871 putR: 1819
%\end{verbatim}
%
%\begin{verbatim}
%eMerge 65432
%getL: 1067 getR: 830
%putL: 655 putR: 410
%mMerge 65432
%getL: 656 getR: 411
%putL: 655 putR: 410
%\end{verbatim}
%
%\begin{verbatim}
%eMerge 76543
%getL: 3261 getR: 735
%putL: 2894 putR: 365
%mMerge 76543
%getL: 2895 getR: 366
%putL: 2894 putR: 365
%\end{verbatim}
%
%\begin{verbatim}
%eMerge 87654
%getL: 1355 getR: 1024
%putL: 844 putR: 509
%mMerge 87654
%getL: 845 getR: 510
%putL: 844 putR: 509
%\end{verbatim}

\section{Related work}

We organize related work in the form of seven ways
to discover Matrix Code:
flowcharts,
automata theory,
abstract state machines,
augmented transition networks,
logic programming,
tail-recursion optimization,
and
recursive program schemes.

\paragraph{Flowcharts}
The following comment has been made on Matrix Code:
``\emph{Although it reeks of flowcharts,
the proposal has some merit}.''
The comment has some merit:
flowcharts are indeed closely related to Matrix Code.
Flowcharts were widely used
as an informal programming notation
from the early 1950s to 1970.
Floyd \cite{fld67} showed
how assertions and verification conditions
can prove a flowchart partially correct.
Hoare \cite{hr69} introduced the notation of triples
for the verification conditions
and cast Floyd's method in the form of inference rules
for control structures such as \\
\verb"      while ... do ...    "
and 
\verb"      if ... then ... else ..."

Dijkstra observed that verifying assertions
are difficult to find for existing code,
so that an attempt at verification is a costly undertaking
with an uncertain outcome. 
He argued \cite{djk68a,djkInfotech71}
that code and correctness should be ``developed in parallel''.
The proposal seems to have found no response,
if only for the lack of specifics in the proposal.

Given the fact that Dijkstra's proposal was considered
unrealistically utopian, and still is,
it is interesting to read
what seems to be the first treatise \cite{gldNmn46}
on programming in the modern sense, published in 1946.
Here programs are expressed in the form of \emph{flow diagrams}.
At first sight one might think
that these are flowcharts under another name.
This is not the case:
flow diagrams consist of executable code
integrated with assertions, with the understanding
that a consistent flow diagram proves the correctness
of the computations performed by it.

The imperative part of a flow diagram was translated
to machine code
(this was before the appearance of assemblers).
I found no indication in \cite{gldNmn46} 
that it was even contemplated to split off
the imperative part of the flow diagram.
Thus we see that what was a vague proposal \cite{djk68a,djkInfotech71},
and regarded as unrealistically utopian in 1970,
was fully worked out in 1946
and may have become a practical reality in 1951
when the IAS machine became operational.

By the time flowcharts appeared,
the proof part of flow diagrams had been dropped.
And apparently forgotten,
for Floyd's discovery was published in 1967
and universally acknowledged as such.
Floyd's format is rather different,
and, in our opinion,
preferable to the flow diagrams of \cite{gldNmn46}.
Matrix Code can be regarded as a simplification
of Floyd's flowchart annotated with assertions,
a simplification made possible by the use of binary relations
that provide a common generalization of statements and tests.
Apt and Schaerf unify statements and tests
in their nondeterministic control structures \cite{ptSchrf97}.

\paragraph{Automata theory}
DSMs can be regarded as a realization
of Dana Scott's idea
\cite{sctt67} to put an end to the proliferation
of new variations of FSM by replacing them by programs
defined to run on suitably defined computers.
DSMs are very different from the programs proposed by
Scott. Scott's programs are unlike FSMs; DSMs closely
resemble FSMs. Paradoxically, DSMs, in the form
of Matrix Code, are of practical use;
Scott's programs are not.

\paragraph{Abstract State Machines}
DSMs can be obtained as a drastic simplification of
ASMs \cite{brgr03} where evolving algebras are
replaced by binary relations over data states and
formulas of logic are replaced by guards.
One might think that guards are a special case
of the formulas of the ASMs.
There is however a fundamental difference: regarded
as logic formulas, guards have free variables;
the formulas of ASMs do not.

\paragraph{Augmented Transition Networks}
In spite of Scott's plea \cite{sctt67}, variants of
FSM continued to appear.
Of special interest in this context are
\emph{labeled transition systems} which are used
to model and verify reactive systems \cite{brktn08}.
Here the set of states is often infinite
and there is typically no halt state.
Such systems are specified by rules of the
form $P \stackrel{A}{\rightarrow} Q$
to indicate the possibility of a transition from
state $P$ to state $Q$ accompanied by action $A$.
Mathematically the rules are viewed as a ternary relation
containing triples consisting of $P$, $A$, and $Q$.
This is of course unobjectionable,
but the alternative view of the rules as constituting
a matrix indexed by states,
containing in this instance $A$ as element indexed
by $P$ and $Q$ has the advantage of connecting
the theory to that of semilinear programming
in the sense of Parker \cite{prkr87}.
Another variant of FSM are the \emph{augmented transition networks}
used in linguistics \cite{wds70}.
The modification of flowcharts by means of binary relations
was introduced in \cite{vnmdn79}.
These can be viewed as augmented transition networks
with binary relations as labels on the transition arrows. 

\paragraph{Logic Programming}
The property that a code matrix is both a set of logical formulas
and an executable program is reminiscent of logic programming,
especially its aspect of separating logic from control
\cite{kwl79a}.
A special form of logic program corresponding to imperative
programs was investigated in \cite{cvn81}.

\paragraph{Recursive program schemes}
De Bakker and de Roever \cite{ddr73}
modeled programming constructs such as if-then-else and while-do.
For both guards and assignments they used binary relations
among what we call data states.

\paragraph{Tail-recursion optimization}
An attractive way of deriving efficient imperative code
is to use a recursive
definition of the function to be computed as starting point.
These can sometimes be transformed to a form
in which there is a single recursive call
and where this call occurs as the last statement
of the function. A further transformation replaces this call
by the more efficient {\tt goto} statement.
The result is similar to the result of translating
a code matrix to executable code.
The definition of the function can then be used to obtain
an assertion verifying the transformed program.
This is used in logic programming \cite{cvn81}.

\section{Conclusions}

In this paper we write programs as matrices with
binary relations as elements.
These matrices can be regarded as transformations
in a generalized vector space,
where vectors have assertions about data states as elements.
Computations of the programs are characterized by
powers of the matrix and verified assertions
show up as generalized eigenvectors of the matrix.
Such results may be dismissed as frivolous theorizing.
It seems to us that they are related to the following practical
benefits.

Our motivation was to address
the fact that imperative programming is in an unsatisfactory
state compared to functional and logic programming.
In the latter paradigms,
implementation is, or is close to, specification.
In imperative programming the relation between implementation
and specification is the verification problem,
a problem considered too hard for the practising programmer.
We proposed Matrix Code as an imperative programming language
where the same construct can be read as logical formula
and can serve as basis for a routine translation to Java,
\verb"C", or \verb"C++".

Matrix Code is only applicable to small algorithms.
Take it as a warning sign when it no longer fits on the back
of an envelope.
Yet it can play a useful role in large programs.
Even the largest software system is ultimately subdivided into
functions or methods.
Software engineering wisdom is unanimous in declaring
any function that is not small as a ``code smell''
and hence a candidate for refactoring.
Everyone of these many small functions
is a candidate for derivation by Matrix Code.

%\vfill\eject
Experience so far suggests that it is possible
to develop algorithms incrementally
by small, obvious steps from the specification.
In this paper we go through such steps for
an algorithm to fill a table with prime numbers
using the method of trial division.
Whether or not this success is an exceptional case,
it seems certain that progress has been made in the direction
of the old dream according to which the production
of verified code is facilitated by developing
proof and code in parallel.  

\section*{Acknowledgments}
Thanks to Paul McJones and Mantis Cheng for their help.
I am grateful to the reviewers for PPDP 2012
and to the ones for \emph{Science of Computer Programming}
for their careful reading
and for their suggestions for improvement.
This research benefited from facilities provided
by the University of Victoria and by the Natural Science
and Engineering Research Council of Canada.


\begin{thebibliography}{10}

\bibitem{ptSchrf97}
K.R. Apt and A. Schaerf.
\newblock Search and imperative programming.
\newblock {\em POPL '97},
pages 67--79.

\bibitem{brktn08}
C. Baier and J.P. Katoen.
\newblock {\em Principles of Model Checking}.
\newblock MIT Press, 2008.

\bibitem{brgr03}
Egon B\"orger and Robert St\"ark.
\newblock {\em Abstract state machines:
   a method for high-level system design and analysis}.
\newblock Springer, 2003.

\bibitem{cvn81}
Keith L. Clark and M.H. van Emden.
\newblock Consequence Verification of Flowcharts.
\newblock {\em IEEE Transactions on Software Engineering},
SE-7:52--60, January 1981.

\bibitem{cnw71}
J.H. Conway.
\newblock {\em Regular Algebra and Finite Machines}.
\newblock Chapman and Hall, 1971.

\bibitem{ddr73}
J.W. de Bakker and W.P. de Roever.
\newblock A Calculus for Recursive Program Schemes.
\newblock {\em Automata, Languages, and Programming},
M. Nivat (ed.), 1973.

\bibitem{djk68a}
E.W. Dijkstra.
\newblock A constructive approach
          to the problem of program correctness.
\newblock {\em BIT}, 8:174--186, 1968.

\bibitem{djkInfotech71}
Edsger~W. Dijkstra.
\newblock Concern for correctness as a guiding principle for program
  composition.
\newblock In J.S.J. Hugo, editor, {\em The Fourth Generation}, pages 359--367.
  Infotech, Ltd, 1971.

\bibitem{djkddh72}
Edsger~W. Dijkstra.
\newblock Notes on structured programming.
\newblock In O.-J. Dahl, E.W. Dijkstra, and C.A.R. Hoare, editors, {\em
  Structured Programming}, pages 1--72. Academic Press, 1972.

%\bibitem{djk76}
%E.W. Dijkstra.
%\newblock {\em A Discipline of Programming}.
%\newblock Prentice Hall, 1976.

\bibitem{fld67}
Robert~W. Floyd.
\newblock Assigning meanings to programs.
\newblock In J.T. Schwartz, editor, {\em Proceedings Symposium
  in Applied Mathematics}, pages 19--32.
  American Mathematical Society, 1967.

\bibitem{gldNmn46}
H.H. Goldstine and J. von Neumann.
\newblock Planning and coding of problems
for an electronic computing instrument. Part II, volume 1, 1946.
\newblock Reprinted in: \emph{John von Neumann: Collected Works},
Pages 80 -- 151, volume V.
\newblock A.H. Taub, editor. Pergamon Press, 1963. 

\bibitem{hr69}
C.A.R. Hoare.
\newblock An axiomatic basis for computer programming.
\newblock {\em Communications of the ACM}, 12(10):576--583, 1969.

\bibitem{kwl79a}
R.A. Kowalski.
\newblock Algorithm = Logic + Control. 
\newblock {\em Comm. ACM}, 22:424--436, 1979.

%\bibitem{lwsppd98}
%H.R. Lewis and C.H. Papadimitriou.
%\newblock \emph{Elements of the Theory of Computation}.
%\newblock Prentice Hall, 1998.

\bibitem{mnsk67}
Marvin L. Minsky.
\newblock \emph{Computation: Finite and Infinite Machines}.
\newblock Prentice Hall, 1967.

\bibitem{prkr87}
D.~Stott Parker.
\newblock Partial order programming.
\newblock Technical Report CSD-870067, Computer Science Department,
  University of California at Los Angeles, 1987.

\bibitem{prr90}
Dominique Perrin.
\newblock Finite Automata.
\newblock Jan van Leeuwen, editor,
{\em Handbook of Theoretical Computer Science, volume B}, pages 1--57.
  Elsevier, 1990.

\bibitem{sctt67}
Dana Scott.
\newblock Some definitional suggestions for automata theory.
\newblock {\em Journal of Computer and Systems Sciences},
1:187--212, 1967.

\bibitem{vnmdn79}
M.H. van Emden.
\newblock Programming with verification conditions.
\newblock {\em IEEE Transactions on Software Engineering},
vol. 3(1979), pp 148--159.

\bibitem{wds70}
W.A. Woods.
\newblock Transition network grammars.
\newblock {\em Comm. ACM}, 13:591--606, 1970.

\end{thebibliography}
\end{document}